\documentclass[
    twocolumn,
	prd,
	amssymb,
	preprintnumbers,superscriptaddress,
	nofootinbib]{revtex4-1}

\pdfoutput=1
\usepackage{amssymb,amsmath,latexsym,mathrsfs}
\usepackage[normalem]{ulem}
\usepackage{url}
\usepackage{hyperref}
\usepackage{enumitem}
\usepackage{graphicx}
\usepackage[usenames,dvipsnames]{xcolor}
\newcommand{\mco}{M}
\newcommand{\msun}{M_\odot}
\newcommand{\fdm}{f_{\rm DM}}

\newcommand{\beq}{\begin{equation}} 
\newcommand{\eeq}{\end{equation}}  
\newcommand{\bea}{\begin{eqnarray}}  
\newcommand{\eea}{\end{eqnarray}}

\newcommand{\Fig}[1]{Fig.~\ref{#1}}

\allowdisplaybreaks

\newcommand{\pL}{\left(} \newcommand{\pR}{\right)}

\begin{document}

\title{Gravitational wave constraints on extended dark matter structures}
\author{Djuna Croon}
\affiliation{Institute for Particle Physics Phenomenology, Department of Physics, Durham University, Durham DH1 3LE, U.K.}
\email{djuna.l.croon@durham.ac.uk}
\author{Seyda Ipek}
\affiliation{Carleton University  1125 Colonel By Drive, Ottawa, Ontario K1S 5B6, Canada}
\email{SeydaIpek@cunet.carleton.ca}
\author{David McKeen}
\affiliation{TRIUMF, 4004 Wesbrook Mall, Vancouver, BC V6T 2A3, Canada}
\email{mckeen@triumf.ca}

\begin{abstract}
We generalise existing constraints on primordial black holes to dark objects with extended sizes using the aLIGO design sensitivity. We show that LIGO is sensitive to dark objects with radius $O(10-10^3~{\rm km})$ if they make up more than $\sim O(10^{-2}-10^{-3})$ of dark matter. 
\end{abstract}
\preprint{IPPP/22/36}

\maketitle

\section{Introduction}
The nature of dark matter (DM) is one of the most important open questions in particle astrophysics. DM candidates span a range of mass scales, from several solar masses to below an eV, with varying types of new interactions. Many popular DM candidates, \emph{e.g.} axions, condense into compact objects, microhalos, or Bose-Einstein condensates. 

Primordial black holes (pBHs) are the most well studied compact objects. 
Strong constraints on pBHs as a DM candidate exist (see, e.g.,~\cite{Green:2020jor}) in the range $ 10^{-11}-10^{1}\msun$ from gravitational microlensing surveys \cite{Alcock_1998,Niikura:2017zjd,Niikura:2019kqi} and in the range $ 10^{1}-3 \times 10^{2}\msun$ from gravitational waves \cite{Ali-Haimoud:2017rtz,Kavanagh:2018ggo,LIGOScientific:2019kan,Chen:2019irf}. Under certain assumptions, these constraints limit pBHs in this mass range to make up a subfraction of DM $ \lesssim 10^{-1}$. Extended objects can in principle be constrained similarly: the gravitational microlensing constraints have recently been generalized to compact objects with extended sizes \cite{Croon:2020wpr,Croon:2020ouk}. It was found that compact objects with a radius smaller than the typical Einstein radius associated with an observation can be constrained, leading to strong constraints across a similar range for DM structures with $ R \lesssim R_\odot$, with the range in masses probed decreasing with increasing radius. 

In this work, we consider the generalisation of gravitational wave sensitivity to extended compact objects (COs). Under the assumption that extended structures form binaries in a similar fashion to pBHs -- a good assumption throughout the inspiral phase of a merger, when the distance between objects is large compared to their radii -- we compute the sensitivity of ground-based experiments LIGO/Virgo/KAGRA (LVK) to gravitational waves from mergers of these objects. We use the frequency associated with the innermost stable circular orbit (ISCO) as an estimate of the peak frequency to which the current experimental searches are sensitive, a methodology suggested in \cite{Giudice:2016zpa}. As the ISCO frequency scales with the compactness $ C = G_N M/R$ as $ f_{\rm ISCO} \propto C^{3/2}$, extended objects have a smaller merger frequency, implying reduced sensitivity in particular at smaller masses (corresponding to gravitational wave events with smaller amplitude).

We consider different merger histories and their impact on the corresponding sensitivities to compact objects. Importantly, compact objects formed in the early Universe, before matter-radiation equality, can be most stringently constrained, but even the most conservative assumptions on the merger history can lead to a constraint on the dark matter fraction $ < 10^{-1}$ for compact objects with masses similar to the LVK black holes and neutron stars. 

\section{Gravitational waves from compact object mergers}
\label{sec:COmergers}
Here we will examine the gravitational wave emission of a binary merger in the inspiral phase. In this phase of the merger, the orbits are circular (gravitational wave emission rapidly circularises an equal mass binary, even if initially eccentric), and the objects are treated like point masses. The waveform is well approximated by the post-Newtonian expansion.

The characteristic strain during the inspiral phase in the quadrupole approximation is given by (see, e.g. \cite{Flanagan:2005yc,Maggiore:2007ulw})
\begin{equation}
\begin{split}
    h_{+}(f) &= \frac{e^{i \Psi_+ (f) }}{\pi^{2/3}} \sqrt{\frac{5}{24}} \frac{1}{d} \frac{(G_N M_c)^{5/6}}{f^{7/6}} 
    \pL \frac{1 + \cos^2 \theta}{2}\pR
    \\ h_{\times}(f) &= \frac{e^{i \Psi_\times (f) }}{\pi^{2/3}} \sqrt{\frac{5}{24}} \frac{1}{d} \frac{(G_N M_c)^{5/6}}{f^{7/6}} \cos \theta \\
    \quad \quad 
    & \text{where} \quad 
    M_c = \frac{(M_1 M_2)^{3/5}}{ (M_1 + M_2)^{1/5}}.
\end{split}
    \label{eq:quadrupolestrain}
\end{equation}
Here $\theta $ is the angle between the normal to the orbit and the line of sight, $d$ is the distance to the source, and the phases $ \Psi_\times = \Psi_+ + \pi /2$ are given by 
\begin{equation}
    \Psi_+ (f) = 2 \pi f (t_c + d) - \Phi_0 - \frac{\pi}{4} + \frac{3}{4} \pL 8 \pi G_N M_c f \pR^{-5/3}
\end{equation}
with $\Phi_0$ the value of the common phase $\Phi$ at the time of coalescence $t_c$.

The assumption of circularity cannot be used for the phase of the binary that ensues after the innermost stable circular orbit (ISCO) has been crossed.
The ISCO can therefore be taken to mark the end of the inspiral phase of a compact object merger. (The true end of the inspiral phase must be determined in numerical simulations, see the discussion in \cite{Giudice:2016zpa,Ajith:2009bn}).
For point-like objects like primordial black holes, $f_{\rm ISCO}$ is given by 
\begin{equation}
    f_{\rm ISCO} = \frac{1}{6^{3/2}\pi M_{\rm tot}}.
\end{equation}
The frequency associated with the generalized ISCO for a compact object with compactness $C=G_N M/R$ is given by
\begin{equation}
\label{eq:fisco}
    f_{\rm ISCO} = \frac{C^{3/2}}{3^{3/2}\pi M_{\rm tot}},
\end{equation}
where $M_{\rm tot} = M_1 + M_2$ is the total mass in the binary. 
For reference, neutron stars have compactness $C \sim 0.1 $.
The approximation in \eqref{eq:fisco} is valid in the non-relativistic limit and breaks down above a certain compactness. For example, this breakdown happens above $ C \gtrsim 0.15$ for a neutron star with polytropic equation of state 
\cite{Lai:1996sv}.  
Staying general about the compact objects and their equation of state, in this work we will use \eqref{eq:fisco},
which may lead to a slight overestimation in sensitivity during the inspiral phase for the most compact objects (close to the PBH curves in Fig.~\ref{fig:fDMplot}). However, this effect may be partially mitigated by the additional gravitational radiation emitted during the merger regime.

The LVK collaboration finds gravitational waves from binary mergers using templates of the inspiral phase. 
We can estimate the sensitivity to extended objects using the signal-to-noise ratio $\rho$:
\[\rho^2 = 4\int_0^{f_{\rm ISCO}}\frac{|h(f)|^2}{h_{\rm exp}(f)}\,df~,\]
with the value averaged over sky position, inclination, and polarisation $\langle \rho^2 \rangle = 4 \rho^2/25$ \cite{Giudice:2016zpa}. We require that $\langle \rho \rangle \geq 8$. This assumes a waveform, which in general differs from the templates used for black holes for compact objects due to tidal effects during the inspiral phase. Such tidal effects cannot be computed in generality as they depend on the detailed equation of state of the object. Here we are mostly interested in $estimating$ the sensitivities of GW detectors, and therefore ignore these effects. 
For $h_{\rm exp}(f)$, we use the aLIGO design sensitivity in~\cite{LIGOsensitivity}, and we will assume equal mass binaries, $M_1 = M_2 = M $ such that $M_{\rm tot} = 2 M$ and $M_c = 2^{-1/5} M$. 
As there is no particular expectation for the angle $\theta$, we average over it in our results.

\begin{figure*}
    \centering
    \includegraphics[width=.99\textwidth]{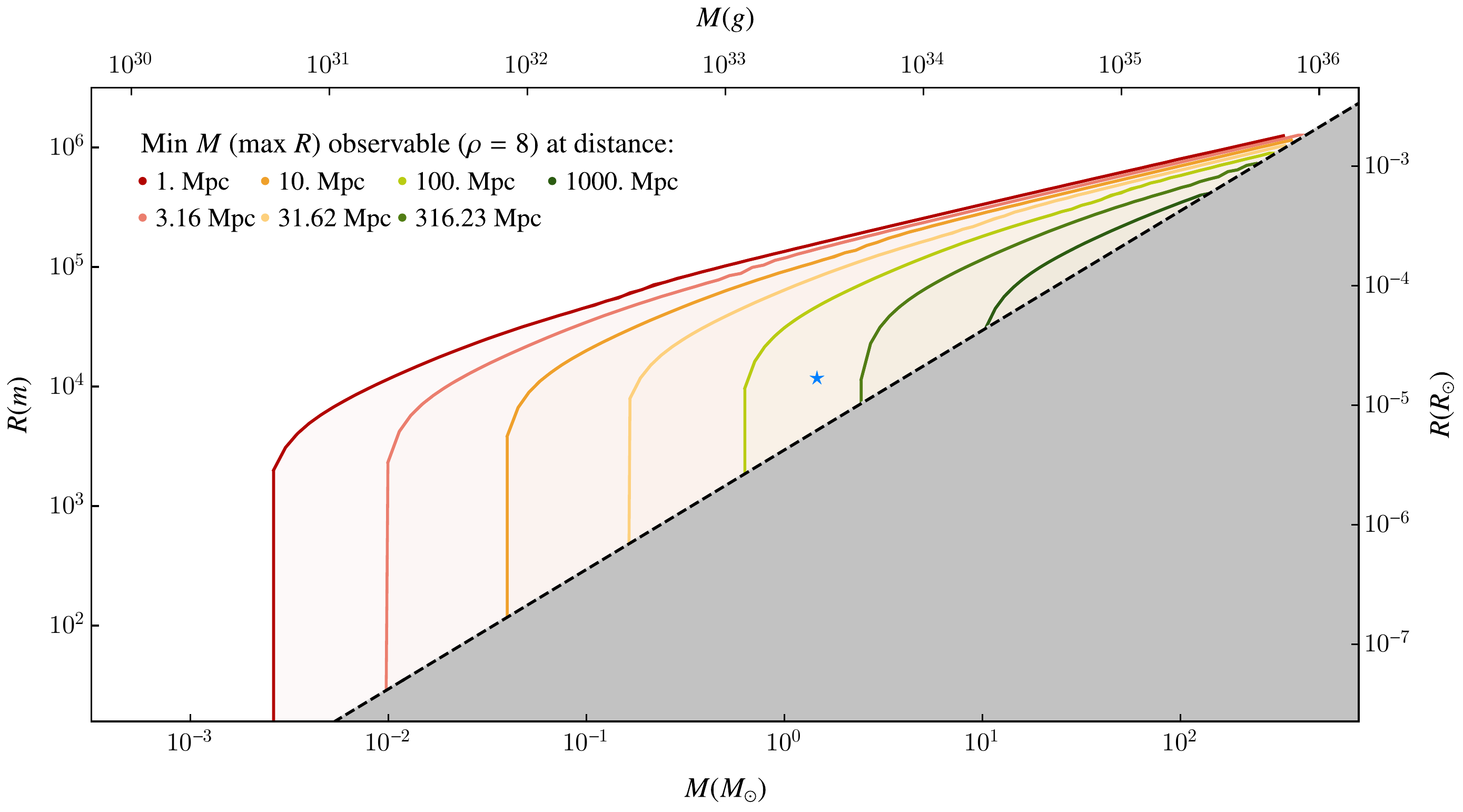}
    \caption{LIGO sensitivity to compact objects during the inspiral regime of a circular equal mass binary.  The regions from left to right indicate observability of mergers at increasing distance. The shaded gray region gives the Schwarzschild radius of a compact object with mass $M$. The star indicates a typical neutron star, for example the primary component of GW170817 \cite{LIGOScientific:2017vwq}, of mass $M=1.46\,  \rm M_\odot$ and radius $R=12 \rm \, km$.}
    \label{fig:sensitivity}
\end{figure*}

We show the result of this computation in Fig.~\ref{fig:sensitivity}. In this plot, the minimal mass (or the maximal radius) of a compact object merger at various distances (increasing from left to right) are shown. The shaded gray area indicates radii below the Schwarzschild radius; on the dashed black line the compact objects are black holes. It is seen that LVK can observe black holes up to masses of a few hundred $M_\odot$. 
The sensitivity drops off for mergers of lighter objects, which can only be observed if they happen at small distances.  
As we will see below, the effective detector volume of the LVK experiment at design sensitivity implies that the larger masses dominate the sensitivity. This is true in particular for objects with large radii. 
Inspirals of objects with radii $> 10^3 $ km, which include double white dwarf binaries, are unlikely to be observed by ground-based interferometers at any mass, due to the small $f_{\rm ISCO} $.

From the plot it is seen that for a certain CO mass, the distance at which the merger is observable decreases as the radius of the object increases. For a given detector volume, this implies a reduced sensitivity to mergers of compact objects with larger radii.

\section{Compact object merger rate}

We assume binaries of compact objects with equal mass $\mco$ that merged recently enough to be observed in presently running experiments. 
We also assume that these binaries formed in similar ways to PBH binaries; in the early universe as well as in present-day halos through gravitational radiation. 
In this section we estimate the merger rates for both of these possibilities. 
The radii of the compact objects to which LVK are sensitive in the inspiral phase does not play a significant role in the formation of the binaries, and therefore the results of this section can be straightforwardly applied to any binary merger in the inspiral phase.

\subsection{Merger rate in present-day halos}

Two compact objects can become gravitationally bound through gravitational radiation. Assuming two equal masses $M$ moving with a relative velocity $v$, the cross-section for this process is~\cite{Mouri:2002mc}
\begin{align}
\begin{split}
    \sigma &= \pi \left(\frac{85\pi}{3}\right)^{2/7}\left(\frac{2G\mco}{c^2}\right)^2 \left(\frac{c}{v}\right)^{18/7}
    \\ &\simeq 4.25\times 10^{-14}\,{\rm pc^2} \left(\frac{\mco}{50\msun}\right)^2\left(\frac{200~{\rm km/s}}{v}\right)^{18/7}\,.
\end{split}
\end{align}
Here we have followed the example of~\cite{Mouri:2002mc} and neglected three-body encounters, which typically lead to wide binaries which do not merge in a Hubble time \cite{Bird:2016dcv}.

The merger rate of two compact objects in a galaxy can be estimated as \cite{physics3020026}
\begin{align}
\Gamma_{\rm per~galaxy} = \frac12 N\,n \,\langle\sigma v\rangle = \frac12 V\, n^2\,\langle\sigma v\rangle ~,
\end{align}
where $N = n V$ is the number of compact objects in the galaxy, $n$ is the number \emph{density} of compact objects and $V$ is the volume of the galaxy. A factor of one half is included to account for double counting.

Making a back-of-the-envelope estimate for this rate is illustrative. Let us assume that all DM in the universe is in  galaxies with mass $M_{\rm gal}$ and uniform mass density $\rho = 0.3\, {\rm GeV/cm^3}\simeq0.08\,\msun/{\rm pc^3}$. If the compact objects make up a fraction $\fdm$ of DM, we have 
\begin{align}
    V = \frac{M_{\rm gal}}{\rho}\,,~~~ n=\fdm\frac{\rho}{\mco}~.
\end{align}
With this the merger rate in such a galaxy becomes
\begin{align}
\begin{split}
    \Gamma_{\rm per~galaxy}\simeq& \frac{1}{2} \frac{M_{\rm gal}}{\rho} \fdm^2\left( \frac{\rho}{\mco} \right)^2\,4.25\times 10^{-14}{\rm pc^2} 
    \\ &\times \left(\frac{\mco}{50\msun}\right)^2\left(\frac{200~{\rm km/s}}{v}\right)^{18/7}v  
    \\
    \simeq& 1.4\times 10^{-11}\fdm^2 \pL \frac{M_{\rm gal}}{10^{12}\msun}\pR
    \\ &\times
    \left(\frac{200~{\rm km/s}}{v}\right)^{11/7}{\rm yr}^{-1}~.
\end{split}
\label{eq:Gammapergalaxy}
\end{align}
Note that this rate does not depend on the mass of the compact object. Assuming all of the matter in the universe is bound in these galaxies, the number density of galaxies with mass $M_{\rm gal}$ in a comoving volume in the universe is 
\begin{align}
    n_{\rm gal} \sim \frac{\rho_{\rm matter}}{M_{\rm gal}}\sim 0.034\,\pL \frac{10^{12}\msun}{M_{\rm gal}}\pR \, {\rm Mpc^{-3}}
\end{align}

Finally, our back-of-the-envelope estimate for rate of mergers per comoving volume in the universe is
\begin{equation}
\begin{split}
    \left(\frac{\Gamma}{V}\right)_{\rm boe} &= n_{\rm gal}\Gamma_{\rm per~galaxy}\\
    &\simeq 4\times 10^{-4}\fdm^2 \, \left(\frac{200~{\rm km/s}}{v}\right)^{11/7} {\rm Gpc^{-3}\,yr^{-1}} ~.
\end{split}
\label{eq:totmergerrate}
\end{equation}
Note that this result does not depend on the mass of the galaxy, nor on the compact object mass, as a result of \eqref{eq:Gammapergalaxy}.

\begin{figure*}
    \centering
    \includegraphics[width=\textwidth]{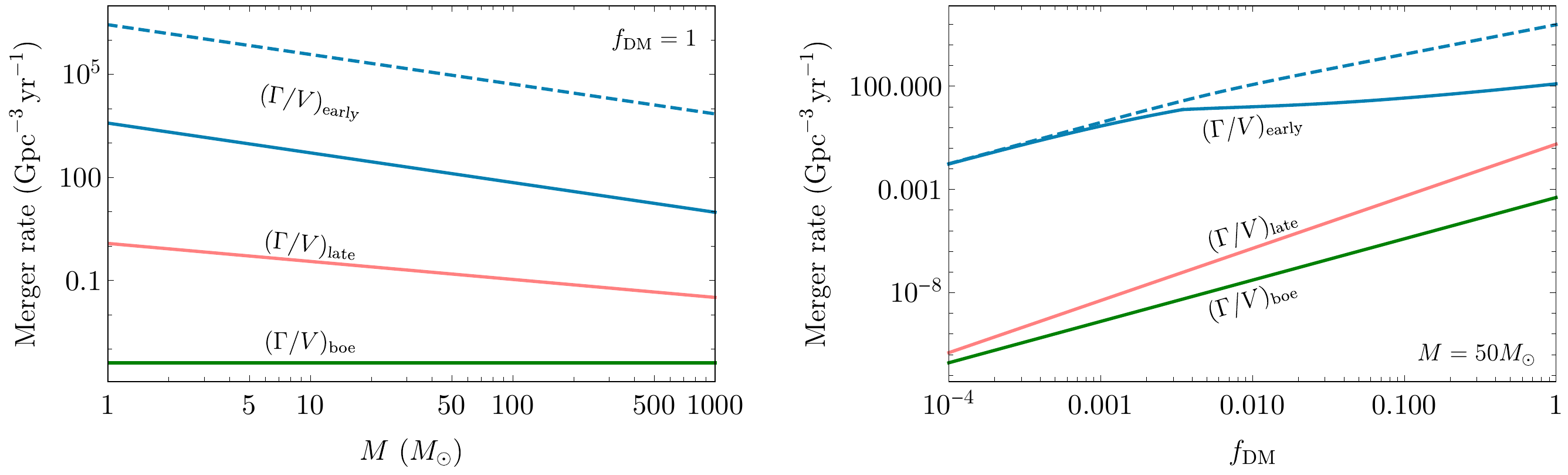}~
    \caption{Comparing the merger rates for binaries that formed in recent galaxies $\Gamma_{\rm late}$, in the early universe $\Gamma_{\rm early}$ and our back-of-the-envelope calculation, $\Gamma_{\rm boe}$. We take $N_c=13$ in \eqref{eq:lessconservative} and $ v = 200 $ km/s in \eqref{eq:totmergerrate}. For $\Gamma_{\rm early}$, the dashed line corresponds to the unsuppressed rate in \eqref{eq:Gearly} and the solid line is the rate with the suppression  factors described in \eqref{eq:Gearly2}.}
    \label{fig:rates}
\end{figure*}

More careful calculations of the merger rate use an Navarro–Frenk–White (NFW) profile for the DM density, a Maxwell-Boltzmann distribution for the DM (and hence compact object) velocity and a halo mass function \cite{Bird:2016dcv}. Most of the common halo mass functions feature an exponential drop for larger galaxies ($M_{\rm gal}\sim 10^{10-15}\msun$). Lower mass galaxies have a smaller velocity dispersion, enhancing the total merger rate per comoving volume. Hence, smaller galaxies dominate the merger rate.\footnote{Jenkins mass function~\cite{Jenkins:2000bv} has an artificial cutoff at $\sim 10^6\msun$. The rate derived from this halo mass function~\cite{Bird:2016dcv} matches our back-of-the-envelope result.} However, one needs to impose a lower limit on the mass of the galaxies because small galaxies can evaporate before creating a merger signal. Ref. \cite{Bird:2016dcv} requires that the smallest halos have at least 13 PBHs. This choice introduces a dependence on the mass of the merging objects, leading to the following merger rate:
\begin{align}
\begin{split}
        \left(\frac{\Gamma}{V}\right)_{\rm late}\simeq & 6.96 \fdm^{53/21} N_c^{-125/84} \label{eq:lessconservative}\\
       &\times \left(\frac{M}{50\msun}\right)^{-11/21}{\rm Gpc^{-3}\,yr^{-1}}~,
\end{split}
    \end{align}
where $N_c = M_{\rm halo}/\mco\geq 13$ for the smallest halos considered to calculate the rate. The cutoff mass increases as $M/\fdm$ and the density-squared in the rate formula brings $\fdm^2$, which gives the $\fdm^{53/21}$ scaling. Using a different DM density profile does not significantly affect the above result 
as long as the density profile does not increase faster than $1/r$ as $r\to 0$, since the only place it comes in to play is the radial integral $\int dr\, r^2 \rho_{\rm DM}^2$.

\subsection{Merger rate in the early universe}
If two compact objects are close enough in the early universe, they decouple from the Hubble flow long before matter-radiation equality and form a binary~\cite{Sasaki:2016jop, Nakamura:1997sm,Ali-Haimoud:2017rtz}. Due to tidal forces generated by the external gravitational potential of other massive objects, the objects can avoid a head-on collision. The spatial and angular distribution of these objects in the early universe, as well as the redshift the binaries are formed, affect the binary formation rate, and hence the present-day merger rate. Furthermore, these probabilities depend on the fraction of dark matter these objects form in a complicated way. For the binaries that formed before matter-radiation equality and merged only recently, the rate is calculated in \cite{Ali-Haimoud:2017rtz} to be
\begin{align}
\begin{split}
        \left(\frac{\Gamma}{V}\right)_{\rm early}^{\rm unsuppressed} &= 9.33\times 10^4 \fdm^2 (\fdm^2+(0.005)^2)^{-21/74}
    \\ &\quad\times\left(\frac{M}{ 50\msun}\right)^{-32/37} {\rm Gpc^{-3}\,yr^{-1}}~. \label{eq:Gearly}
    \end{split}
\end{align}
Here the number $3.4\times 10^{-5}$ is the variance of density perturbations of the rest of the dark matter at matter-radiation equality. If $\fdm \lesssim 0.005$, the rate scales as $\fdm^2$, while it scales as $\fdm^{53/37}$ for $\fdm \gtrsim 0.005$. 
The unusual power law associated with $M$ and $\fdm$ arises from the relationship between the angular momentum of the binaries and their lifetime as well as finding the most probable separation between the compact objects that results in mergers today. 

\begin{figure*}
    \centering
    \includegraphics[width=.48\textwidth]{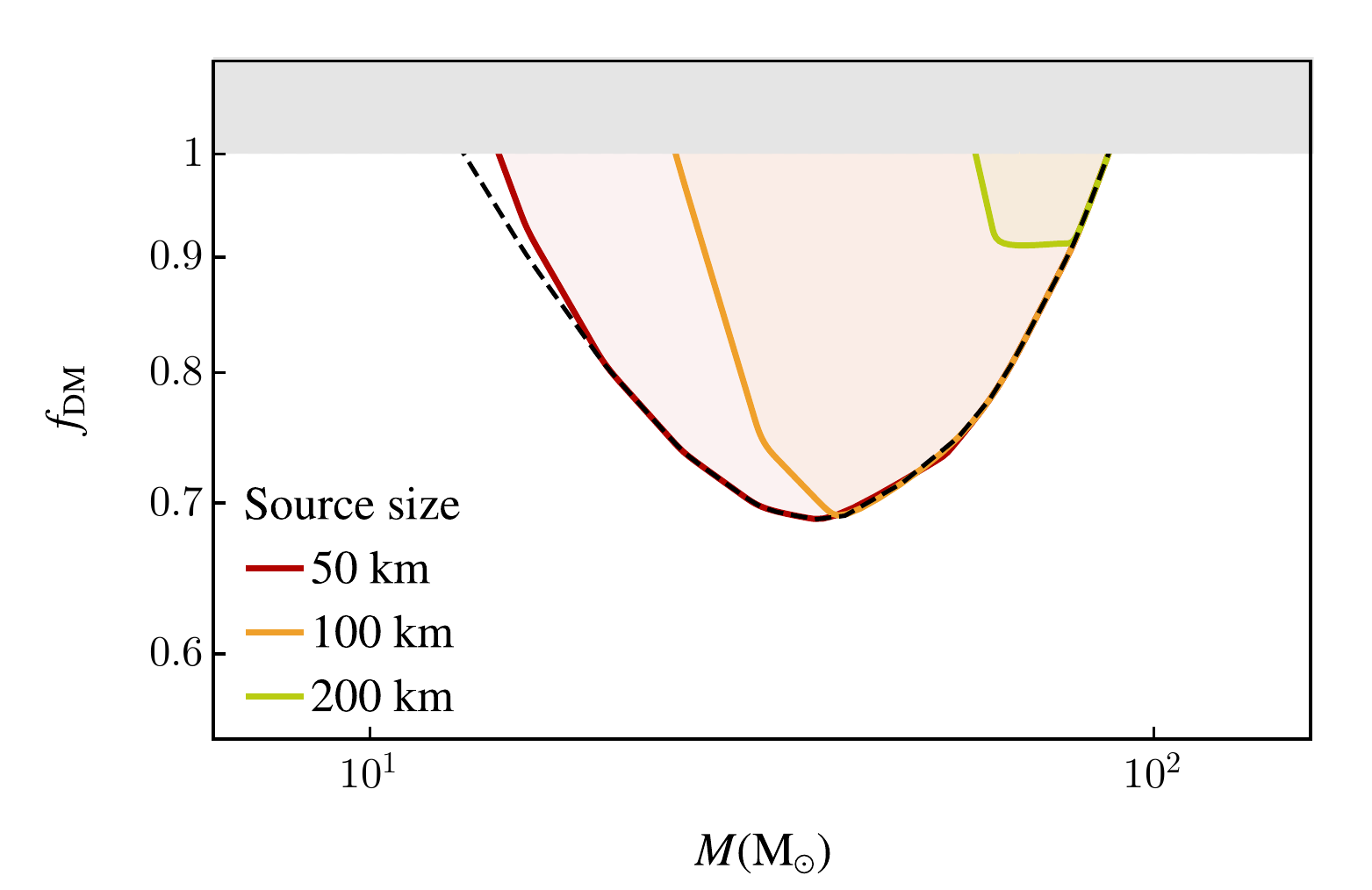}
     \includegraphics[width=.48 \textwidth]{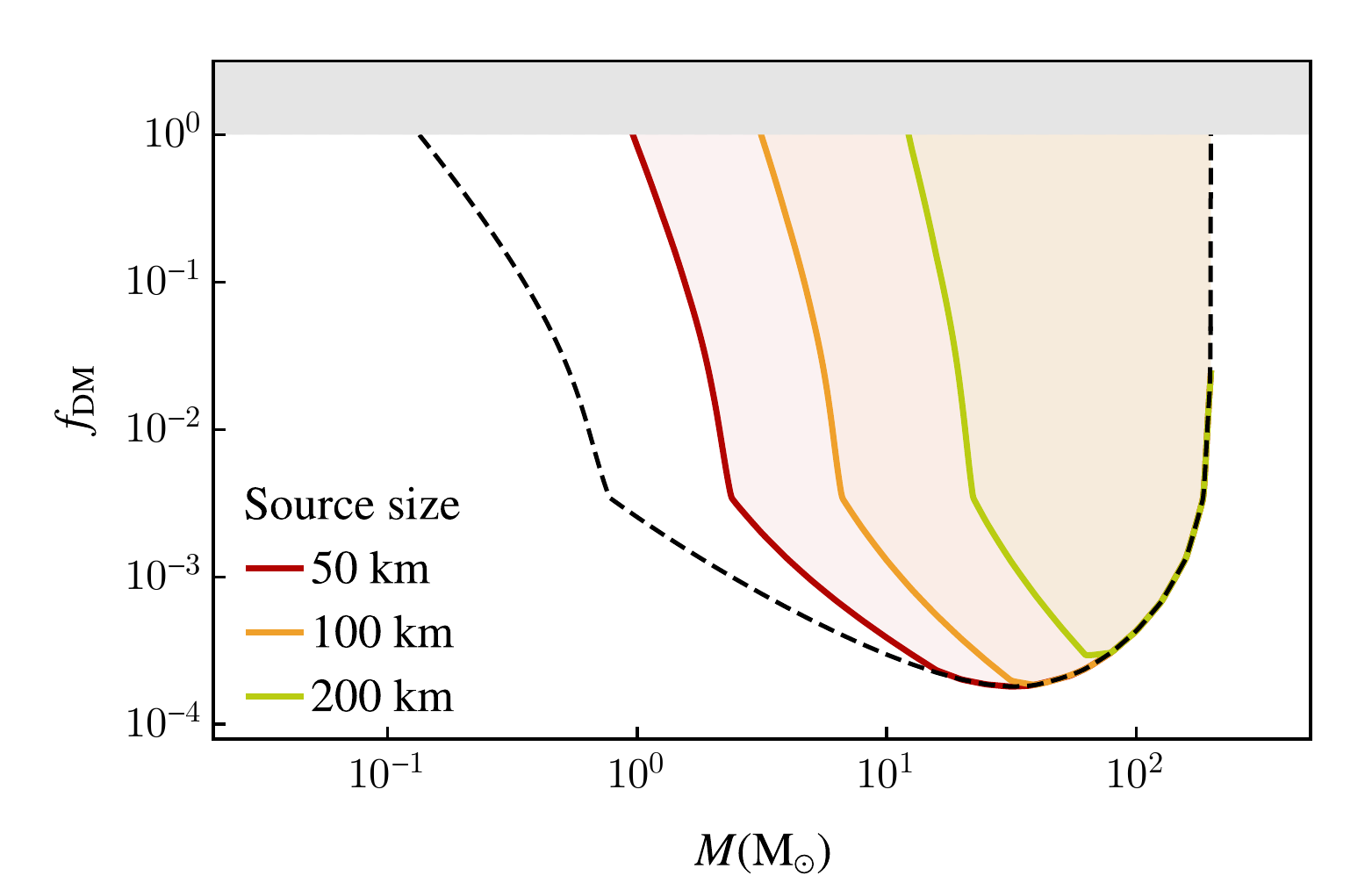}
    \caption{Sensitivity to $f_{\rm DM}$ for compact objects of various sizes. \textbf{(Left)} For binaries forming in the present-day halos using the more optimistic estimate \eqref{eq:lessconservative}, with the more conservative estimate \eqref{eq:totmergerrate} no sensitivity is found. \textbf{(Right)} For binaries forming in the early Universe using the rate with suppression effects described in \eqref{eq:Gearly2}. The dashed black line indicates the PBH constraint for which the radius of the object is the Swarzschild radius.   }
    \label{fig:fDMplot}
\end{figure*}

Although analytical arguments that lead to the above merger rate suggest that possible later interactions with other black holes and baryon accretion do not alter the structure of binaries merging today~\cite{Ali-Haimoud:2017rtz}, there are N-body simulations showing noticeable effects due to nearby black holes and other matter~\cite{Raidal:2018bbj, Jedamzik:2020ypm}. These effects can significantly suppress the binary merger rate for $\fdm\gtrsim10^{-2}$. Detailed calculations can be found in~\cite{Vaskonen:2019jpv, Raidal:2018bbj}. Although we use the approximation given in Appendix A in \cite{Hutsi:2020sol} in our numerical calculations, we find that the following formula is close to the more detailed results for $\fdm \gtrsim 10^{-5}$ and is more illuminating for the reader:

\begin{align}
\label{eq:Gearly2}
\begin{split}
    \left(\frac{\Gamma}{V}\right)_{\rm early}^{\rm suppressed} \simeq& 1.81\times 10^{4} \fdm^{53/37}\left(\frac{M}{50\msun}\right)^{-32/37} 
\\ & \times
S_{\rm \rm early}S_{\rm late}~{\rm Gpc}^{-3}{\rm yr}^{-1}\,, 
\end{split}
\end{align}
where
\begin{align}
S_{\rm early} &= {\rm min}\left(\left(\frac{\fdm}{0.004}\right)^{-1/14}, \left(\frac{\fdm}{0.004}\right)^{1/3}\right)\,, \\ 
S_{\rm late}&={\rm min}\left(1,~9.6\times 10^{-3}\fdm^{-0.65}e^{0.03\ln^2\fdm}\right)\,. \notag
\end{align}

We show these rates in \Fig{fig:rates}. The early-universe binary formation rate is much larger than the present-day rate, even when suppression due to interactions with nearby objects are included. We take the early universe rate as the most optimistic scenario for observations and the present-day binary merger rate as a conservative scenario.

\section{aLIGO sensitivity to compact objects}

In the left panel of Fig.~\ref{fig:fDMplot} we show a sensitivity to the sub-fraction of dark matter compact objects can comprise, based on  \eqref{eq:lessconservative}.
This sensitivity estimate is based on  circular, equal mass binaries in the inspiral phase under the assumptions discussed in Section \ref{sec:COmergers}.
It is seen that LIGO is as sensitive to compact objects with radii $\lesssim 50$ km as it is to primordial black holes, whereas the sensitivity quickly reduces for larger radii, in particular at smaller CO masses.
This is consistent with the expectation based on the scaling of the ISCO frequency with the compactness $f_{\rm ISCO} \propto C^{3/2}$.
For celestial objects with small radii, the sensitivity to compact object mergers largely compliments the constraints found from microlensing surveys, which extend to $ \sim 10 \, \msun$ \cite{Croon:2020ouk}.

In the right panel of Fig.~\ref{fig:fDMplot} we show the equivalent sensitivity assuming that the compact objects formed in the early Universe (before matter-radiation equality), leading to the much larger merger rate  \eqref{eq:Gearly2}.
In addition, the difference in scaling with $M$ of this rate implies a better sensitivity to lower mass objects. Importantly, for radii $ < 50 $ km this includes sub-solar mass objects, which can be straightforwardly distinguished from neutron stars by their mass. Including microlensing constraints \cite{Croon:2020ouk}, for radii $< 100 $ km the dark matter subfraction can be constrained to below $ f_{\rm DM }<10^{-1}$ for $ 10^{-10}-10^2 M_\odot$.

\section{Discussion}
In this work, we have demonstrated the sensitivity of the ground-based interferometer network to compact object mergers. We have focussed on the aLIGO design sensitivity and shown that with a conservative merger rate of 1/2 years the dark matter fraction comprising compact objects with radius $< \mathcal{O}(10^3~{\rm km})$ can be constrained during the inspiral regime of the merger. COs with larger radii merge at lower frequencies, but can potentially be constrained with upcoming space-based or atomic interferometers. 

The derived constraints sensitively depend on the present-day CO merger rate, itself a function of the CO history. 
Two COs can become gravitationally bound inside present-day halos due to gravitational wave emission, but this is  subdominant compared to binaries of COs which were formed close together in the early Universe. Although the former formation channel can already be used to exclude $f_{\rm DM}$ over several decades in CO mass, information about the formation may significantly improve the sensitivity. 

Importantly, CO binaries originating in the early Universe have a merger rate which scales non-linearly with $ f_{\rm DM}^2$. We show the sensitivity corresponding to \eqref{eq:Gearly2} in the left panel of Fig.~\ref{fig:fDMplot}. In present-day halos, the merger rate depends on the low-end tail of the chosen halo mass function. In the right panel of Fig.~\ref{fig:fDMplot} we show the estimated sensitivity using the merger rate given \eqref{eq:lessconservative} for binaries formed in the present-day halos.

We have made a few simplifying assumptions in this work. Firstly, we have assumed equal-mass binaries. We expect that without a-priori merger preference of particular mass ratios, relaxing this assumption will lead to minor differences. Secondly, we have not considered binaries of a CO and a neutron star or black hole. Taking such mergers into account would lead to stronger sensitivity, though the partial overlap of halo mass profiles of DM and luminous matter renders such mergers subdominant. Specific clustering of COs (as recently discussed for PBHs in \cite{Pilipenko:2022emp}) may also increase the merger rate and therefore modify our results. 
Moreover, we have only considered the inspiral regime of compact object mergers. The merger phase will lead to additional gravitational wave radiation, such that our results can be seen as conservative. 

Individual CO mergers may be distinguished from binary black hole or neutron star mergers through their mass, absence of electromagnetic signal, or tidal distortions, impacting the gravitational waveform via the tidal love number~\cite{Postnikov:2010yn,Cardoso:2017cfl,Sennett:2017etc,Palenzuela:2017kcg,Nelson:2018xtr,Bezares:2018qwa,Cardoso:2019rvt,Ciancarella:2020msu,Poisson:2020vap,DeLuca:2021ite,Helfer:2021brt,Collier:2022cpr,Karkevandi:2021ygv,RafieiKarkevandi:2021hcc}. The latter is a particularly important observation channel for the CO with radii $> 50 \rm\, km $ considered in this work, although some subtleties may arise in its interpretation~\cite{Gralla:2017djj,Chia:2020yla,Creci:2021rkz}. Populations of CO may be distinguished via population properties such as mass, spin, and eccentricity distributions. 

Future experiments may improve on the sensitivity presented here both by increased experimental sensitivity and longer observation times. The next generation of ground-based interferometers (including the Einstein Telescope \cite{Gupta:2022qgg}) may be an order of magnitude more sensitive in characteristic strain, implying an increased sensitivity to $f_{\rm DM}$ of a factor of three or more, depending on the dominant binary formation mechanism. 

\section*{Acknowledgement}
SI thanks Nicholas Orlofsky for discussions on black hole mergers. DC is supported by the STFC under Grant No. ST/T001011/1.
This work is supported in part by the Natural Sciences and Engineering Research Council (NSERC) of Canada.
TRIUMF receives federal funding via a contribution agreement with the National Research Council (NRC) of Canada.
This research was carried out in part at the Munich Institute for Astro- and Particle Physics (MIAPP) which is funded by the Deutsche Forschungsgemeinschaft (DFG, German Research Foundation) under Germany´s Excellence Strategy – EXC-2094 – 390783311.

\bibliographystyle{JHEP}
\bibliography{ref}

\providecommand{\href}[2]{#2}\begingroup\raggedright\begin{thebibliography}{10}

\bibitem{Green:2020jor}
A.M.~Green and B.J.~Kavanagh, \emph{{Primordial Black Holes as a dark matter
  candidate}}, \href{https://doi.org/10.1088/1361-6471/abc534}{\emph{J. Phys.
  G} {\bfseries 48} (2021) 043001}
  [\href{https://arxiv.org/abs/2007.10722}{{\ttfamily 2007.10722}}].

\bibitem{Alcock_1998}
C.~Alcock, R.A.~Allsman, D.~Alves, R.~Ansari, {\'{E}}.~Aubourg, T.S.~Axelrod
  et~al., \emph{{EROS} and {MACHO} combined limits on planetary-mass dark
  matter in the galactic halo}, \href{https://doi.org/10.1086/311355}{\emph{The
  Astrophysical Journal} {\bfseries 499} (1998) L9}.

\bibitem{Niikura:2017zjd}
H.~Niikura et~al., \emph{{Microlensing constraints on primordial black holes
  with Subaru/HSC Andromeda observations}},
  \href{https://doi.org/10.1038/s41550-019-0723-1}{\emph{Nature Astron.}
  {\bfseries 3} (2019) 524} [\href{https://arxiv.org/abs/1701.02151}{{\ttfamily
  1701.02151}}].

\bibitem{Niikura:2019kqi}
H.~Niikura, M.~Takada, S.~Yokoyama, T.~Sumi and S.~Masaki, \emph{{Constraints
  on Earth-mass primordial black holes from OGLE 5-year microlensing events}},
  \href{https://doi.org/10.1103/PhysRevD.99.083503}{\emph{Phys. Rev. D}
  {\bfseries 99} (2019) 083503}
  [\href{https://arxiv.org/abs/1901.07120}{{\ttfamily 1901.07120}}].

\bibitem{Ali-Haimoud:2017rtz}
Y.~Ali-Ha\"\i{}moud, E.D.~Kovetz and M.~Kamionkowski, \emph{{Merger rate of
  primordial black-hole binaries}},
  \href{https://doi.org/10.1103/PhysRevD.96.123523}{\emph{Phys. Rev. D}
  {\bfseries 96} (2017) 123523}
  [\href{https://arxiv.org/abs/1709.06576}{{\ttfamily 1709.06576}}].

\bibitem{Kavanagh:2018ggo}
B.J.~Kavanagh, D.~Gaggero and G.~Bertone, \emph{{Merger rate of a subdominant
  population of primordial black holes}},
  \href{https://doi.org/10.1103/PhysRevD.98.023536}{\emph{Phys. Rev. D}
  {\bfseries 98} (2018) 023536}
  [\href{https://arxiv.org/abs/1805.09034}{{\ttfamily 1805.09034}}].

\bibitem{LIGOScientific:2019kan}
{\scshape LIGO Scientific, Virgo} collaboration, \emph{{Search for Subsolar
  Mass Ultracompact Binaries in Advanced LIGO\textquoteright{}s Second
  Observing Run}},
  \href{https://doi.org/10.1103/PhysRevLett.123.161102}{\emph{Phys. Rev. Lett.}
  {\bfseries 123} (2019) 161102}
  [\href{https://arxiv.org/abs/1904.08976}{{\ttfamily 1904.08976}}].

\bibitem{Chen:2019irf}
Z.-C.~Chen and Q.-G.~Huang, \emph{{Distinguishing Primordial Black Holes from
  Astrophysical Black Holes by Einstein Telescope and Cosmic Explorer}},
  \href{https://doi.org/10.1088/1475-7516/2020/08/039}{\emph{JCAP} {\bfseries
  08} (2020) 039} [\href{https://arxiv.org/abs/1904.02396}{{\ttfamily
  1904.02396}}].

\bibitem{Croon:2020wpr}
D.~Croon, D.~McKeen and N.~Raj, \emph{{Gravitational microlensing by dark
  matter in extended structures}},
  \href{https://doi.org/10.1103/PhysRevD.101.083013}{\emph{Phys. Rev. D}
  {\bfseries 101} (2020) 083013}
  [\href{https://arxiv.org/abs/2002.08962}{{\ttfamily 2002.08962}}].

\bibitem{Croon:2020ouk}
D.~Croon, D.~McKeen, N.~Raj and Z.~Wang, \emph{{Subaru-HSC through a different
  lens: Microlensing by extended dark matter structures}},
  \href{https://doi.org/10.1103/PhysRevD.102.083021}{\emph{Phys. Rev. D}
  {\bfseries 102} (2020) 083021}
  [\href{https://arxiv.org/abs/2007.12697}{{\ttfamily 2007.12697}}].

\bibitem{Giudice:2016zpa}
G.F.~Giudice, M.~McCullough and A.~Urbano, \emph{{Hunting for Dark Particles
  with Gravitational Waves}},
  \href{https://doi.org/10.1088/1475-7516/2016/10/001}{\emph{JCAP} {\bfseries
  10} (2016) 001} [\href{https://arxiv.org/abs/1605.01209}{{\ttfamily
  1605.01209}}].

\bibitem{Flanagan:2005yc}
E.E.~Flanagan and S.A.~Hughes, \emph{{The Basics of gravitational wave
  theory}}, \href{https://doi.org/10.1088/1367-2630/7/1/204}{\emph{New J.
  Phys.} {\bfseries 7} (2005) 204}
  [\href{https://arxiv.org/abs/gr-qc/0501041}{{\ttfamily gr-qc/0501041}}].

\bibitem{Maggiore:2007ulw}
M.~Maggiore, \emph{{Gravitational Waves. Vol. 1: Theory and Experiments}},
  Oxford Master Series in Physics, Oxford University Press (2007).

\bibitem{Ajith:2009bn}
P.~Ajith et~al., \emph{{Inspiral-merger-ringdown waveforms for black-hole
  binaries with non-precessing spins}},
  \href{https://doi.org/10.1103/PhysRevLett.106.241101}{\emph{Phys. Rev. Lett.}
  {\bfseries 106} (2011) 241101}
  [\href{https://arxiv.org/abs/0909.2867}{{\ttfamily 0909.2867}}].

\bibitem{Lai:1996sv}
D.~Lai and A.G.~Wiseman, \emph{{Innermost stable circular orbit of inspiraling
  neutron star binaries: Tidal effects, postNewtonian effects and the neutron
  star equation of state}},
  \href{https://doi.org/10.1103/PhysRevD.54.3958}{\emph{Phys. Rev. D}
  {\bfseries 54} (1996) 3958}
  [\href{https://arxiv.org/abs/gr-qc/9609014}{{\ttfamily gr-qc/9609014}}].

\bibitem{LIGOsensitivity}
``{Updated Advanced LIGO sensitivity design curve}.''
  \url{https://dcc.ligo.org/LIGO-T1800044/public}.

\bibitem{LIGOScientific:2017vwq}
{\scshape LIGO Scientific, Virgo} collaboration, \emph{{GW170817: Observation
  of Gravitational Waves from a Binary Neutron Star Inspiral}},
  \href{https://doi.org/10.1103/PhysRevLett.119.161101}{\emph{Phys. Rev. Lett.}
  {\bfseries 119} (2017) 161101}
  [\href{https://arxiv.org/abs/1710.05832}{{\ttfamily 1710.05832}}].

\bibitem{Mouri:2002mc}
H.~Mouri and Y.~Taniguchi, \emph{{Runaway merging of black holes: analytical
  constraint on the timescale}},
  \href{https://doi.org/10.1086/339472}{\emph{Astrophys. J. Lett.} {\bfseries
  566} (2002) L17} [\href{https://arxiv.org/abs/astro-ph/0201102}{{\ttfamily
  astro-ph/0201102}}].

\bibitem{Bird:2016dcv}
S.~Bird, I.~Cholis, J.B.~Mu\~noz, Y.~Ali-Ha\"\i{}moud, M.~Kamionkowski,
  E.D.~Kovetz et~al., \emph{{Did LIGO detect dark matter?}},
  \href{https://doi.org/10.1103/PhysRevLett.116.201301}{\emph{Phys. Rev. Lett.}
  {\bfseries 116} (2016) 201301}
  [\href{https://arxiv.org/abs/1603.00464}{{\ttfamily 1603.00464}}].

\bibitem{physics3020026}
V.D.~Stasenko and A.A.~Kirillov, \emph{The merger rate of black holes in a
  primordial black hole cluster},
  \href{https://doi.org/10.3390/physics3020026}{\emph{Physics} {\bfseries 3}
  (2021) 372}.

\bibitem{Jenkins:2000bv}
A.~Jenkins, C.S.~Frenk, S.D.M.~White, J.M.~Colberg, S.~Cole, A.E.~Evrard
  et~al., \emph{{The Mass function of dark matter halos}},
  \href{https://doi.org/10.1046/j.1365-8711.2001.04029.x}{\emph{Mon. Not. Roy.
  Astron. Soc.} {\bfseries 321} (2001) 372}
  [\href{https://arxiv.org/abs/astro-ph/0005260}{{\ttfamily
  astro-ph/0005260}}].

\bibitem{Sasaki:2016jop}
M.~Sasaki, T.~Suyama, T.~Tanaka and S.~Yokoyama, \emph{{Primordial Black Hole
  Scenario for the Gravitational-Wave Event GW150914}},
  \href{https://doi.org/10.1103/PhysRevLett.117.061101}{\emph{Phys. Rev. Lett.}
  {\bfseries 117} (2016) 061101}
  [\href{https://arxiv.org/abs/1603.08338}{{\ttfamily 1603.08338}}].

\bibitem{Nakamura:1997sm}
T.~Nakamura, M.~Sasaki, T.~Tanaka and K.S.~Thorne, \emph{{Gravitational waves
  from coalescing black hole MACHO binaries}},
  \href{https://doi.org/10.1086/310886}{\emph{Astrophys. J. Lett.} {\bfseries
  487} (1997) L139} [\href{https://arxiv.org/abs/astro-ph/9708060}{{\ttfamily
  astro-ph/9708060}}].

\bibitem{Raidal:2018bbj}
M.~Raidal, C.~Spethmann, V.~Vaskonen and H.~Veerm\"ae, \emph{{Formation and
  Evolution of Primordial Black Hole Binaries in the Early Universe}},
  \href{https://doi.org/10.1088/1475-7516/2019/02/018}{\emph{JCAP} {\bfseries
  02} (2019) 018} [\href{https://arxiv.org/abs/1812.01930}{{\ttfamily
  1812.01930}}].

\bibitem{Jedamzik:2020ypm}
K.~Jedamzik, \emph{{Primordial Black Hole Dark Matter and the LIGO/Virgo
  observations}},
  \href{https://doi.org/10.1088/1475-7516/2020/09/022}{\emph{JCAP} {\bfseries
  09} (2020) 022} [\href{https://arxiv.org/abs/2006.11172}{{\ttfamily
  2006.11172}}].

\bibitem{Vaskonen:2019jpv}
V.~Vaskonen and H.~Veerm\"ae, \emph{{Lower bound on the primordial black hole
  merger rate}}, \href{https://doi.org/10.1103/PhysRevD.101.043015}{\emph{Phys.
  Rev. D} {\bfseries 101} (2020) 043015}
  [\href{https://arxiv.org/abs/1908.09752}{{\ttfamily 1908.09752}}].

\bibitem{Hutsi:2020sol}
G.~H\"utsi, M.~Raidal, V.~Vaskonen and H.~Veerm\"ae, \emph{{Two populations of
  LIGO-Virgo black holes}},
  \href{https://doi.org/10.1088/1475-7516/2021/03/068}{\emph{JCAP} {\bfseries
  03} (2021) 068} [\href{https://arxiv.org/abs/2012.02786}{{\ttfamily
  2012.02786}}].

\bibitem{Pilipenko:2022emp}
S.~Pilipenko, M.~Tkachev and P.~Ivanov, \emph{{The evolution of a primordial
  binary black hole due to interaction with cold dark matter and the formation
  rate of gravitational wave events}},
  \href{https://arxiv.org/abs/2205.10792}{{\ttfamily 2205.10792}}.

\bibitem{Postnikov:2010yn}
S.~Postnikov, M.~Prakash and J.M.~Lattimer, \emph{{Tidal Love Numbers of
  Neutron and Self-Bound Quark Stars}},
  \href{https://doi.org/10.1103/PhysRevD.82.024016}{\emph{Phys. Rev. D}
  {\bfseries 82} (2010) 024016}
  [\href{https://arxiv.org/abs/1004.5098}{{\ttfamily 1004.5098}}].

\bibitem{Cardoso:2017cfl}
V.~Cardoso, E.~Franzin, A.~Maselli, P.~Pani and G.~Raposo, \emph{{Testing
  strong-field gravity with tidal Love numbers}},
  \href{https://doi.org/10.1103/PhysRevD.95.084014}{\emph{Phys. Rev. D}
  {\bfseries 95} (2017) 084014}
  [\href{https://arxiv.org/abs/1701.01116}{{\ttfamily 1701.01116}}].

\bibitem{Sennett:2017etc}
N.~Sennett, T.~Hinderer, J.~Steinhoff, A.~Buonanno and S.~Ossokine,
  \emph{{Distinguishing Boson Stars from Black Holes and Neutron Stars from
  Tidal Interactions in Inspiraling Binary Systems}},
  \href{https://doi.org/10.1103/PhysRevD.96.024002}{\emph{Phys. Rev. D}
  {\bfseries 96} (2017) 024002}
  [\href{https://arxiv.org/abs/1704.08651}{{\ttfamily 1704.08651}}].

\bibitem{Palenzuela:2017kcg}
C.~Palenzuela, P.~Pani, M.~Bezares, V.~Cardoso, L.~Lehner and S.~Liebling,
  \emph{{Gravitational Wave Signatures of Highly Compact Boson Star Binaries}},
  \href{https://doi.org/10.1103/PhysRevD.96.104058}{\emph{Phys. Rev. D}
  {\bfseries 96} (2017) 104058}
  [\href{https://arxiv.org/abs/1710.09432}{{\ttfamily 1710.09432}}].

\bibitem{Nelson:2018xtr}
A.~Nelson, S.~Reddy and D.~Zhou, \emph{{Dark halos around neutron stars and
  gravitational waves}},
  \href{https://doi.org/10.1088/1475-7516/2019/07/012}{\emph{JCAP} {\bfseries
  1907} (2019) 012} [\href{https://arxiv.org/abs/1803.03266}{{\ttfamily
  1803.03266}}].

\bibitem{Bezares:2018qwa}
M.~Bezares and C.~Palenzuela, \emph{{Gravitational Waves from Dark Boson Star
  binary mergers}},
  \href{https://doi.org/10.1088/1361-6382/aae87c}{\emph{Class. Quant. Grav.}
  {\bfseries 35} (2018) 234002}
  [\href{https://arxiv.org/abs/1808.10732}{{\ttfamily 1808.10732}}].

\bibitem{Cardoso:2019rvt}
V.~Cardoso and P.~Pani, \emph{{Testing the nature of dark compact objects: a
  status report}},
  \href{https://doi.org/10.1007/s41114-019-0020-4}{\emph{Living Rev. Rel.}
  {\bfseries 22} (2019) 4} [\href{https://arxiv.org/abs/1904.05363}{{\ttfamily
  1904.05363}}].

\bibitem{Ciancarella:2020msu}
R.~Ciancarella, F.~Pannarale, A.~Addazi and A.~Marciano, \emph{{Constraining
  mirror dark matter inside neutron stars}},
  \href{https://doi.org/10.1016/j.dark.2021.100796}{\emph{Phys. Dark Univ.}
  {\bfseries 32} (2021) 100796}
  [\href{https://arxiv.org/abs/2010.12904}{{\ttfamily 2010.12904}}].

\bibitem{Poisson:2020vap}
E.~Poisson, \emph{{Compact body in a tidal environment: New types of
  relativistic Love numbers, and a post-Newtonian operational definition for
  tidally induced multipole moments}},
  \href{https://doi.org/10.1103/PhysRevD.103.064023}{\emph{Phys. Rev. D}
  {\bfseries 103} (2021) 064023}
  [\href{https://arxiv.org/abs/2012.10184}{{\ttfamily 2012.10184}}].

\bibitem{DeLuca:2021ite}
V.~De~Luca and P.~Pani, \emph{{Tidal deformability of dressed black holes and
  tests of ultralight bosons in extended mass ranges}},
  \href{https://doi.org/10.1088/1475-7516/2021/08/032}{\emph{JCAP} {\bfseries
  08} (2021) 032} [\href{https://arxiv.org/abs/2106.14428}{{\ttfamily
  2106.14428}}].

\bibitem{Helfer:2021brt}
T.~Helfer, U.~Sperhake, R.~Croft, M.~Radia, B.-X.~Ge and E.A.~Lim,
  \emph{{Malaise and remedy of binary boson-star initial data}},
  \href{https://doi.org/10.1088/1361-6382/ac53b7}{\emph{Class. Quant. Grav.}
  {\bfseries 39} (2022) 074001}
  [\href{https://arxiv.org/abs/2108.11995}{{\ttfamily 2108.11995}}].

\bibitem{Collier:2022cpr}
M.~Collier, D.~Croon and R.K.~Leane, \emph{{Tidal Love Numbers of Novel and
  Admixed Celestial Objects}},
  \href{https://arxiv.org/abs/2205.15337}{{\ttfamily 2205.15337}}.

\bibitem{Karkevandi:2021ygv}
D.R.~Karkevandi, S.~Shakeri, V.~Sagun and O.~Ivanytskyi, \emph{{Bosonic dark
  matter in neutron stars and its effect on gravitational wave signal}},
  \href{https://doi.org/10.1103/PhysRevD.105.023001}{\emph{Phys. Rev. D}
  {\bfseries 105} (2022) 023001}
  [\href{https://arxiv.org/abs/2109.03801}{{\ttfamily 2109.03801}}].

\bibitem{RafieiKarkevandi:2021hcc}
D.~Rafiei~Karkevandi, S.~Shakeri, V.~Sagun and O.~Ivanytskyi, \emph{{Tidal
  Deformability as a Probe of Dark Matter in Neutron Stars}},  in \emph{{16th
  Marcel Grossmann Meeting on~Recent Developments in Theoretical and
  Experimental General Relativity, Astrophysics and Relativistic Field
  Theories}}, 12, 2021 [\href{https://arxiv.org/abs/2112.14231}{{\ttfamily
  2112.14231}}].

\bibitem{Gralla:2017djj}
S.E.~Gralla, \emph{{On the Ambiguity in Relativistic Tidal Deformability}},
  \href{https://doi.org/10.1088/1361-6382/aab186}{\emph{Class. Quant. Grav.}
  {\bfseries 35} (2018) 085002}
  [\href{https://arxiv.org/abs/1710.11096}{{\ttfamily 1710.11096}}].

\bibitem{Chia:2020yla}
H.S.~Chia, \emph{{Tidal deformation and dissipation of rotating black holes}},
  \href{https://doi.org/10.1103/PhysRevD.104.024013}{\emph{Phys. Rev. D}
  {\bfseries 104} (2021) 024013}
  [\href{https://arxiv.org/abs/2010.07300}{{\ttfamily 2010.07300}}].

\bibitem{Creci:2021rkz}
G.~Creci, T.~Hinderer and J.~Steinhoff, \emph{{Tidal response from scattering
  and the role of analytic continuation}},
  \href{https://doi.org/10.1103/PhysRevD.104.124061}{\emph{Phys. Rev. D}
  {\bfseries 104} (2021) 124061}
  [\href{https://arxiv.org/abs/2108.03385}{{\ttfamily 2108.03385}}].

\bibitem{Gupta:2022qgg}
P.K.~Gupta, A.~Puecher, P.T.H.~Pang, J.~Janquart, G.~Koekoek and C.~Broeck
  Van~Den, \emph{{Determining the equation of state of neutron stars with
  Einstein Telescope using tidal effects and r-mode excitations from a
  population of binary inspirals}},
  \href{https://arxiv.org/abs/2205.01182}{{\ttfamily 2205.01182}}.

\end{thebibliography}\endgroup

\end{document}